\documentclass[twocolumn,showpacs,preprintnumbers,floatfix,amsmath,amssymb,superscriptaddress]{revtex4}

\usepackage{graphicx}
\usepackage{dcolumn}
\usepackage{bm}

\begin{document}

\title{Numerical treatment of long-range Coulomb potential with Berggren bases}
\author{N.~Michel}
\affiliation{Department of Physics, Post Office Box 35 (YFL), University of Jyv{\"a}skyl{\"a}, FI-40014 Jyv{\"a}skyl{\"a}, Finland}

\begin{abstract}
The Schr{\"o}dinger equation incorporating the long-range Coulomb potential takes the form of a Fredholm equation whose kernel is singular on its diagonal
when represented by a basis bearing a continuum of states, such as in a Fourier-Bessel transform.
Several methods have been devised to tackle this difficulty, from simply
 removing the infinite-range of the Coulomb potential with a screening or cut function
to using discretizing schemes which take advantage of the integrable character of Coulomb kernel singularities.
However, they have never been tested in the context of Berggren bases, which  allow many-body nuclear wave functions to be expanded,
with halo or resonant properties within a shell model framework.
It is thus the object of this paper to test different discretization schemes
of the Coulomb potential kernel in the framework of complex-energy nuclear physics. 
For that, the Berggren basis expansion of proton states pertaining to the $sd$-shell arising in the $A \sim 20$ region, being typically resonant, will be effected.
Apart from standard frameworks involving a cut function or analytical integration of singularities, a new method will be presented, which replaces diagonal singularities by finite off-diagonal terms.
It will be shown that this methodology surpasses in precision the two former techniques.
\end{abstract}


\bigskip

\maketitle

\section{Introduction} \label{introduction} 
The infinite-range of the Coulomb potential has always been a source of numerous problems in both theoretical and numerical studies of the quantum mechanics of charged particles.

From a theoretical point of view, the $1/r$ Coulomb asymptotics implies
the presence of an essential singularity of the Coulomb Green's function at $k=0$ in the complex plane,
contrary to that of potentials quickly vanishing at $r \rightarrow +\infty$ \cite{Newton_book}.
This has prevented for a long time the use of the simple and elegant demonstration of the completeness of bound and scattering states solutions of the one-body Schr{\"o}dinger equation of R.~Newton \cite{Newton},
as the latter explicitly demands $r^2~|V(r)|$ to be integrable on $[0:+\infty[$, where $V(r)$ is the basis-generating potential. Until recently, no simple proof existed for the Coulomb case, as the only available methods
requested abstract Lebesgue measure theory \cite{Dunford_Schwarz}, thus hindering any physical understanding of this fundamental property. 
Hence, new methodologies had to be devised in order to have both simple and rigorous demonstrations,
based on a generalization of the method of R.~Newton \cite{Akin} or on
closeness arguments with standard complete sets of states \cite{JMP},
which rely heavily on the analytical properties of Coulomb wave functions.

At the numerical level, the computational effort necessary to evaluate Coulomb wave functions is tremendous compared to that of its uncharged counterpart, the analytical Ricatti-Bessel function.
Indeed, even though analytically expressible in terms of confluent hypergeometric functions \cite{Abramowitz_Stegun}, Coulomb wave functions are very difficult to calculate, especially in the complex plane \cite{Thompson,cwfcomplex}.
They can vary by many orders of magnitude over a relatively small region
and thus demand the use of many different numerical techniques \cite{Thompson,cwfcomplex}.
Added to that, they exhibit a cut in the complex plane on $]-\infty:0]$, which creates a supplementary difficulty as the functions issued from continued fractions, for example, are analytical therein \cite{Thompson,cwfcomplex}.
When one considers a Fourier-Bessel transform of potentials bearing a Coulomb tail, no problem appears at the theoretical level.
This comes from the fact that the latter can always be decomposed as $V_0(r) + c/r$, where $V_0(r)$ is a quickly vanishing potential and $c$ is a constant. 
Indeed, on the one hand, the Fourier-Bessel transform of $V_0(r)$ is well-behaved,
and on the other hand, the term proportional to $1/r$ has a Fourier-Bessel transform analytical for all orbital momentum $\ell$, equal to $Q_\ell[(k^2 + {k'}^2)/(2 k k')]/\pi$, where $Q_\ell(x)$
is the Legendre function of the second kind \cite{Abramowitz_Stegun}.
However, $Q_\ell(x)$ diverges like $\ln(1-x)$ when $x \rightarrow 1$ \cite{PRD_sub}, so that the momentum space representation of $V(r)$ has a logarithmic singularity on its diagonal
at $k = k'$, and consequently cannot be discretized readily, as the diagonal of the discretized matrix is infinite. 
Nevertheless, the Fourier-Bessel transform of $1/r$ is integrable, so that it is possible to solve this problem with subtraction frameworks,
where singularities are integrated analytically, leaving a regular rest integral to be handled numerically \cite{NR}. 
This has been successfully applied for the diagonalization of the Coulomb potential in momentum space in Refs.\cite{PRD_sub,PRC_sub,PRA_sub}. 

However, no attempt has been made using Berggren bases \cite{Berggren}.
The latter are sets of wave functions generated by finite-range potentials and comprise bound, resonant, and scattering states of complex-energy, which are solutions of the one-body Schr{\"o}dinger equation. 
Berggren sets of eigenstates are widely used in nuclear physics to solve the many-body problem, 
as they allow complex nuclear wave functions to be expanded in a basis of Slater determinants built from Berggren one-body states, in the so-called Gamow shell model framework (see Ref.\cite{Mic09} for a recent review on that subject).
Due to the presence of scattering states in Berggren bases, the asymptotic behavior of complex many-body wave functions can be precisely reconstructed,
which is impossible to attain in practice with bases of well-bound states, such as sets of harmonic oscillator states.
They are then very well suited to expand halo and unbound many-body nuclear states \cite{Bet02,Bet03,Mic02,Mic03,Mic04,Gaute_Morten,Gaute_Morten_2,Gaute_Morten_3,Mic07,Mic07a,Mic10}, 
whose properties cannot be understood without a precise reproduction of their asymptotic properties.
Generating potentials are usually of Woods-Saxon type, which mimic the effect of the inert core on valence nucleons \cite{WS}.
Hartree-Fock potentials are also utilized, as their variational
character allows configuration mixing to be minimized \cite{Mic04,Gaute_Morten_3}. 
In all these cases, wave functions cannot be conveyed in an analytical form and have to be computed numerically.
Completeness properties of Berggren sets of states have been established firstly by T.~Berggren for uncharged particles \cite{Berggren}, while the charged case could be handled only recently in Ref.\cite{JMP}.
Completeness relations have been tested by numerically diagonalizing Hamiltonians sustaining spherical and axially-deformed nuclear potentials
for both charged and uncharged particles
\cite{Mic03,Mic04,Cocoyoc,Hagen_Vaagen}, but where the Coulomb potential is absent from the diagonalized kernel. 
If one considers nuclear wave functions of a fixed nucleus, this problem is just anecdotic, as, due to its one-body character, the infinite-range part of the Coulomb potential, which is spherical, can always be included
in the basis-generating potential, leaving a finite-range residual interaction to diagonalize \cite{Mic10}. However, this is no longer possible if one wants to calculate observables
connecting nuclei bearing different numbers of protons, as the same Berggren basis has to be used to expand the many-body wave functions of both father and daughter nuclei. 
Problems also occur to calculate isospin operators expectation values with Berggren bases 
(see for example Ref.\cite{Mic10} for a study of isospin mixture of isobaric analog states of $A=6$ nuclei). Indeed, it is needed therein to evaluate one-body matrix elements $\langle a | t^\pm | b  \rangle$,
where $|a \rangle$ and $|b \rangle$ can be scattering wave functions. If proton and neutron wave functions of a given partial wave are generated by the same potential, $\langle a | t^\pm | b  \rangle$ matrix elements 
are proportional to a Dirac delta distribution. 
In the opposite situation, however, Dirac delta distributions cannot arise as proton and neutron radial wave functions of same quantum numbers $(\ell,j)$ are not orthogonal, so that matrix elements are undefined.
Consequently, both proton and neutron Berggren set of states have to be generated by the same one-body Hamiltonian, for example with the basis-generating potential of neutron states, 
so that the infinite-range Coulomb Hamiltonian has to be diagonalized along with the residual nuclear interaction.

Thus, in this paper, a one-body Hamiltonian possessing an infinite-range proton potential, 
thus presenting a Coulomb point potential asymptotic, will be diagonalized with Berggren bases employing three different discretization schemes. 
The first scheme is simply to cut the Coulomb potential at a finite distance $R$, determined so as to provide optimal results. The second scheme is a subtraction technique, akin to that expressed in Ref.\cite{PRA_sub},
and the third scheme consists of replacing the diverging diagonal
elements by off-diagonal terms, which become diagonal at the continuum limit. These methods will be described in detail in Sec.~\ref{theory},
where they will be respectively labeled by ``the cut method,'' the ``subtraction method,'' and the ``off-diagonal method.'' 
Numerical applications will be presented in Sec.~\ref{examples} for three different partial waves, $s_{1/2}$, $d_{3/2}$ and $d_{5/2}$, where both narrow and broad resonant states will be considered.
The considered expanded one-body wave functions, namely the $1s_{1/2}$, $0d_{3/2}$ and $0d_{5/2}$ proton states, 
 arise typically for nuclei for which $A \sim 20$, so that their study is important for future calculations of many-body states of light nuclei with the Gamow shell model.
Numerical methods proper to the use of proton Berggren bases states, generated by a potential with a Coulomb asymptotic, will also be discussed.
The conclusion is stated in Sec.~\ref{conclusion}.

\section{Theoretical background} \label{theory}

\subsection{Expression of the Hamiltonian kernel in a Berggren basis} \label{theory_Berggren_basis}
One considers a Berggren basis of a partial wave of quantum numbers $(\ell,j)$, whose completeness relation reads \cite{Berggren}:
\begin{eqnarray}
\sum_{n \in (b,d)} u_n(r) u_n(r') + \int_{L^+} u_k(r) u_k(r')~dk = \delta(r-r'), \label{Berggren_completeness_relation}
\end{eqnarray}
where $L^+$ is a complex contour of momenta $k$ starting at $k=0$ and ending at $k = +\infty$,
representing the scattering part of the Berggren basis constituted by the $| u_k \rangle$ states,
and where the discrete sum runs over bound $(b)$ and resonant, or decaying $(d)$ states $| u_n \rangle$, 
the latter resonant states being situated between the real $k$-axis and $L^+$ (see Ref.\cite{Mic03} for details).
The energy of the discrete states $| u_n \rangle$ will be denoted as $e_n$ and that of scattering states $| u_k \rangle$ as $e_k$.
Discrete states $| u_n \rangle$ are normalized to one and scattering states $| u_k \rangle$ to Dirac delta \cite{Mic03}.
They are generated by a one-body Hamiltonian of the form:
\begin{eqnarray}
&&h = \frac{\hat{p}^2}{2m} + V_{WS}(r) + V_c(Z_c,r) \label{Hamiltonian} \\
&&V_{WS}(r) = -V_o f(r) - 4 (\vec{\ell} \cdot \vec{s}) V_{so} \frac{1}{r} \left| \frac{df}{dr'} \right|_{r'=r} \nonumber \\
&&f(r) = \left[ 1 + \exp \left( \frac{r-R_0}{d} \right) \right]^{-1} \label{WS_potential} \\
&&V_c(Z_c,r) = \frac{C_c~Z_c~\mbox{erf}(\alpha r)}{r} \label{Coulomb_potential},
\end{eqnarray}
where $m$ is the effective mass of the proton, $V_{WS}(r)$ is a potential of Woods-Saxon type, with its diffuseness $d$, radius $R_0$, central and spin-orbit depths $V_o$ and $V_{so}$, respectively,
$\vec{\ell}$ is the orbital momentum operator and $\vec{s}$ the spin operator,
and $V_c$ is the Coulomb potential, proportional to an error function, defined with the Coulomb constant $C_c$, the charge acting on the proton $Z_c$ and a constant $\alpha$.
The radial function entering the Coulomb potential $V_c(Z_c,r)$ is standard and arises from the use of a Gaussian charge density for the closed core \cite{Sai77,Myo98,idb08}.
For simplicity, the finite-range part of both basis and diagonalized Hamiltonians will be taken to be the same so that they will differ only through their charge $Z_c$.
We will denote the charge of the basis potential as $Z_c^{(b)}$, that of the diagonalized potential $Z_c^{(d)}$, and their difference, which will enter the kernel to diagonalize,
will be denominated as $\Delta Z_c = Z_c^{(d)} - Z_c^{(b)}$. The eigenstate $| \phi \rangle$ of the diagonalized Hamiltonian is to be expanded with Eq.~(\ref{Berggren_completeness_relation}):
\begin{eqnarray}
| \phi \rangle = \sum_{n \in (b,d)} c_n | u_n \rangle + \int_{L^+} c_k | u_k \rangle~dk \label{phi_expansion},
\end{eqnarray}
where the coefficients $c_n$ and $c_k$ have to be determined. Its energy will be designated as $E$.
Using Eqs.~(\ref{Berggren_completeness_relation}) and (\ref{Hamiltonian}), the Fredholm equations which must be solved to ascertain the $c_n$ and $c_k$ coefficients are readily inferred:
\begin{eqnarray}
c_n~e_n &+& \sum_{n' \in (b,d)} c_{n'} \langle u_{n'} | V_c(\Delta Z_c,r) | u_n \rangle \nonumber \\
&+& \int_{L^+} c_{k'} \langle u_{k'} | V_c(\Delta Z_c,r) | u_n \rangle~dk' \nonumber \\
&=& E~c_n \mbox{ } \forall n \in (b,d), \nonumber \\
c_k~e_k &+& \sum_{n' \in (b,d)} c_{n'} \langle u_{n'} | V_c(\Delta Z_c,r) | u_k \rangle  \nonumber \\
&+& \int_{L^+} c_{k'} \langle u_{k'} | V_c(\Delta Z_c,r) | u_k \rangle~dk'  \nonumber \\
&=& E~c_k \mbox{ } \forall k \in L^+ \label{Fredholm_equations}.
\end{eqnarray}
Matrix elements of Eq.~(\ref{Fredholm_equations}) have to be calculated
using a complex rotation of $r$ \cite{Gya71,Simon} due to the unbound character of basis states:
\begin{eqnarray}
&&\langle u_a | V_c(\Delta Z_c,r) | u_b \rangle = \int_{0}^{R} u_a(r) V_c(\Delta Z_c,r) u_b(r)~dr \nonumber \\
&+& \sum_{\stackrel{\omega_a = \pm}{\omega_b = \pm}} e^{i \theta} \int_{0}^{+\infty} u_a^{\omega_a}(z(x)) \frac{C_c~\Delta Z_c}{R + x e^{i \theta}} u_b^{\omega_b}(z(x))~dx \label{Vc_complex_scaling_OBME},
\end{eqnarray}
where $| u_a \rangle$ and $| u_b \rangle$ are two Berggren basis states, $R$ is a radius after which complex rotation is effected, sufficiently large to sustain erf$(\alpha r) \simeq 1$ and $f(r) \simeq 0$ 
in Eqs.~(\ref{WS_potential}) and (\ref{Coulomb_potential}), $\omega = \pm$ characterizes the asymptotic character of wave functions components, as $u(r) = u^+(r) + u^-(r) = C^+ H^+_{\ell \eta}(kr) + C^- H^-_{\ell \eta}(kr)$ for $r>R$,
$C^\pm$ being a constant and $H^\pm_{\ell \eta}(z)$ a Coulomb wave function of orbital momentum $\ell$ and Sommerfeld parameter $\eta$, of outgoing $(+)$ or incoming $(-)$ character,
$\theta$ is the rotation angle, which depends on $\omega_a$ and
$\omega_b$ and is chosen so that the associated improper integral converges and $z(x) = R + x e^{i \theta}$.
Coulomb wave functions are implemented using recently developed techniques mixing analytical formulas and direct integration \cite{cwfcomplex}.
Complex rotation of $r$ is also employed to normalize the discrete states $| u_n \rangle$ [see Eq.~(\ref{Berggren_completeness_relation})] \cite{Mic03}.
One has to pay attention if $k(R + x e^{i \theta})$ crosses the negative real axis with complex rotation of $r$, as $H^\pm_{\ell \eta}(z)$ becomes discontinuous therein.
Indeed, Coulomb wave functions bear a cut by definition, whereas the wave functions $u^\pm(r)$ are everywhere continuous, as solutions of a differential equation of second order.
Crossing cannot occur in practice for resonant states, sustaining sufficiently small width-to-energy ratios, 
while non-crossing can be enforced for scattering states by taking a $L^+$ contour sufficiently close to the real $k$-axis. 
It is, however, unavoidable for bound states when $\cos(\theta) < 0$. 
In the latter case, a linear combination of $H^+_{\ell \eta}(kr)$ and $H^-_{\ell \eta}(kr)$, whose coefficients are simple functions of $\eta$ and $k$ \cite{cwfcomplex},
has to be added to the initial wave function $u(r) = u^+(r) = C^+ H^+_{\ell \eta}(kr)$ to suppress discontinuity.
Because, for bound states, $|H^-_{\ell \eta}(k(R + x e^{i \theta}))| \rightarrow 0$ when $x \rightarrow +\infty$ with $\cos(\theta) < 0$,
this modification does not change the asymptotic properties of $u(r)$.

Due to analyticity of $u_a(r)$ and $u_b(r)$ in the complex $r$-space, the matrix element embedded in Eq.~(\ref{Vc_complex_scaling_OBME}) is independent of $R$ and $\theta$,
so that it is a genuine definition of operators acting on Berggren basis states. However, as one might expect, Eq.~(\ref{Vc_complex_scaling_OBME}) cannot be used for $| u_a \rangle = | u_b \rangle = |u_k \rangle$.
Although improper integrals for which $\omega_a \omega_b = 1$ are always well defined with complex rotation of $r$,
no rotation angle $\theta$ can be found to have improper integrals converged when $\omega_a \omega_b = -1$. 
This is straightforward to demonstrate using the asymptotical expressions of $u^\pm(z)$ for $|z| \rightarrow +\infty$, which reads:
\begin{eqnarray}
u^{\pm}(z) \sim \exp[\pm i(k z - \eta \ln (2 k z))] \label{asymptotic},
\end{eqnarray}
up to an unimportant constant. As a consequence, similarly to the representation of $1/r$ in momentum space, diagonal matrix elements
are infinite  (i.e.~they cannot be regularized with complex rotation of $r$) and off-diagonal matrix elements $\langle u_k | V_c(\Delta Z_c,r) | u_{k'} \rangle$ for which $k \neq k'$, where complex rotation is always available,
diverge like $\ln(k-k')$ when $k' \rightarrow k$.
Modification of diagonal matrix elements in Eq.~(\ref{Vc_complex_scaling_OBME}) is then necessary to discretize Eq.~(\ref{Fredholm_equations}).

\subsection{Discretization of Berggren basis} \label{Berggren_basis_discretization}
The most efficient method to discretize Eq.~(\ref{Fredholm_equations}) is to utilize a Gauss-Legendre quadrature for the $L^+$ contour of Eq.~(\ref{Berggren_completeness_relation}) \cite{NR}.
In order to obtain a symmetric matrix in the resulting eigenproblem, discretized scattering states are set equal to $\sqrt{w_i}~u_{k_i}(r)$,
where $w_i$ is the weight associated to the $k_i$ abscissa. The discretized Berggren completeness relation then reads:
\begin{eqnarray}
&&\sum_{n \in (b,d)} u_n(r) u_n(r') \nonumber \\
&&+ \sum_{i=0}^{N_{GL}-1} u_{k_i}(r) u_{k_i}(r')~w_i \simeq \delta(r-r') \nonumber \\
&&\Leftrightarrow \sum_{i=0}^{N_{res}-1} u_i(r) u_i(r') \nonumber \\
&&+ \sum_{i=N_{res}}^{N-1} u_{k_{i_s}}(r) u_{k_{i_s}}(r')~w_{i_s} \simeq \delta(r-r') \nonumber \\
&&\Leftrightarrow \sum_{i=0}^{N-1} u_i(r) u_i(r') \simeq \delta(r-r') \label{discretized_Berggren_completeness_relation},
\end{eqnarray}
where $N_{res}$ is the number of bound or resonant states, $N_{GL}$  is the number of Gauss-Legendre points, $N = N_{res} + N_{GL}$ is the total number of basis states,
$i_s = i-N_{res}$, notation which we will maintain from now on, and
$u_i(r)$ is a bound or resonant state $u_n(r)$ if $i=n \in [0:N_{res}-1]$ and the discretized state $\sqrt{w_{i_s}}~u_{k_{i_s}}(r)$ if $i \in [N_{res}:N-1]$.
The energy of the $i$-th state will be designated as $e_i$.
The resulting discretized eigenstate implemented from diagonalization of discretized kernel will be naturally denoted as:
\begin{eqnarray}
| \phi \rangle \simeq \sum_{i=0}^{N-1} c_i | u_i \rangle. \label{discretized_phi_expansion}
\end{eqnarray}

\subsection{Cut method} \label{cut_method}
This approximation is the crudest, as it simply removes all improper integrals in Eq.~(\ref{Vc_complex_scaling_OBME}), 
which is equivalent to replacing $V_c(Z_c,r)$ of Eq.~(\ref{Coulomb_potential}) by $V_c(Z_c,r)~He(R-r)$, where $He$ is the Heaviside function. 
As a consequence, Eq.~(\ref{Vc_complex_scaling_OBME}) becomes:
\begin{eqnarray}
\langle u_a | V_c(\Delta Z_c,r) | u_b \rangle = \int_{0}^{R} u_a(r) V_c(\Delta Z_c,r) u_b(r)~dr \label{cut_Coulomb_OBME}
\end{eqnarray}
This removes all singularities in the Hamiltonian kernel of Eq.~(\ref{Hamiltonian}), but at the price of introducing a cut-dependence on $R$ which has to be assessed.
Its discretized eigenproblem is then straightforward to convey utilizing Eq.~(\ref{discretized_phi_expansion}) (see Sec.~\ref{Berggren_basis_discretization} for notations):
\begin{eqnarray}
&&c_i~e_i + \sum_{i'=0}^{N-1} c_{i'} \langle u_{i'} | V_c(\Delta Z_c,r) | u_i \rangle  = E~c_i, \nonumber \\ 
&&0 \leq i < N \label{cut_discretized_Fredholm_equations}.
\end{eqnarray}

\subsection{Subtraction method} \label{subtraction_method}
This is the standard method to treat integrable singularities in Fredholm kernels \cite{NR}. The main idea is to separate the kernel integral into two parts, 
one whose integrand is regular and the other singular but analytically integrable. 
However, contrary to the situation of Ref.\cite{PRA_sub}, it is necessary to terminate the $L^+$ contour at finite $k = k_{max}$.
Indeed, allowing the $L^+$ contour to go to $+\infty$ in the $k$-plane, with Gauss-Laguerre quadrature for example, 
would demand the consideration of wave functions of very high linear momenta, consequently tremendously oscillating.
The latter would have to be used in radial integrals which have to be implemented numerically, such as the nonanalytical kernels of Eq.~(\ref{Fredholm_equations}),
hence inducing numerical instability. Because, in practice, convergence with $k_{max}$ is rather fast, it is preferrable to utilize Gauss-Legendre quadrature on a $L^+$ contour ending at $k = k_{max}$.

In order to apply the subtraction method, the integral of Eq.~(\ref{Fredholm_equations}) involving $k$ and $k'$ is rewritten the following way:
\begin{eqnarray} 
&& \int_{L^+} c_{k'} \langle u_{k'} | V_c(\Delta Z_c,r) | u_k \rangle~dk' \nonumber \\
&=& \int_{L^+} \left[ \int_{0}^{+\infty} \left( c_{k'}~u_{k'}(r)~V_c(\Delta Z_c,r)~u_k(r) \right. \right. \nonumber \\
&-& \left. \left. c_k~s_{k'}(r)~\frac{C_c~\Delta Z_c}{r}~s_k(r) \right) dr \right] dk' \nonumber \\
&+& c_{k} \int_{L^+} \left\langle s_{k'} \left| \frac{C_c~\Delta Z_c}{r} \right| s_k \right\rangle~dk' \label{subtraction_integral},
\end{eqnarray}
where $| s_k \rangle$ is a sine function [i.e.~$\langle r | s_k \rangle = \sqrt{2/ \pi} \sin(kr)$], so that one has just added and subtracted the $\ell = 0$ Fourier-Bessel transform of the Coulomb potential $(C_c~\Delta Z_c)/r$.
The calculation of the radial integrals inside the first $k'$-integral of the right-hand side of Eq.~(\ref{subtraction_integral}) can always be performed with complex rotation of $r$ when $k \neq k'$ 
(see Sec.~\ref{theory_Berggren_basis}).
When $k' \sim k$, the consideration of the asymptotics of $u^\pm_k(r)$ for $r \rightarrow +\infty$ [see Eq.~(\ref{asymptotic})], 
the fact that $2 \pi C^+ C^- = 1$ (arising from Dirac delta
normalization of scattering states \cite{Mic03}) 
and that $c_{k'} - c_k \sim (k'-k) (\partial c_{k''}/\partial k'')_{k'' = k}$ ($c_k$ is an analytic function of $k$) 
imply that the radial integrals problematic with complex rotation of $r$; 
that is those for which $k = k'$ and $\omega_a \omega_b = -1$ (see Sec.~\ref{theory_Berggren_basis} for notations), are convergent in Eq.~(\ref{subtraction_integral}).

The second integral of the right-hand side of Eq.~(\ref{subtraction_integral}) can be evaluated analytically:
\begin{eqnarray}
&&\int_{L^+} \left\langle s_{k'} \left| \frac{C_c~\Delta Z_c}{r} \right| s_k \right \rangle~dk'  \nonumber \\
&=& \frac{C_c~\Delta Z_c}{\pi} \left[ \int_{0}^{k} [ \ln(k + k') - \ln(k - k')]~dk' \right. \nonumber \\
&+& \left. \int_{k}^{k_{max}} [\ln(k + k') - \ln(k' - k)]~dk' \right] \nonumber \\
&=& \frac{C_c~\Delta Z_c}{\pi} \left[ (k_{max}+k) \ln(k_{max}+k) \right. \nonumber \\
&-& \left. (k_{max}-k) \ln(k_{max}-k) - 2 k \ln(k) \right] \label{analytical_integral},
\end{eqnarray}
where the fact that the complex linear momentum $k$ belongs to the $L^+$ contour has been taken into account.
The discretization of Eq.~(\ref{Fredholm_equations}) can now be effected, which results in a different treatment of diagonal and off-diagonal terms involving scattering states 
(see Sec.~\ref{Berggren_basis_discretization} for notations):
\begin{eqnarray}
&&\!\!\!\!\!c_i~e_i + \sum_{i'=0}^{N-1} c_{i'} \langle u_{i'} | V_c(\Delta Z_c,r) | u_i \rangle  = E~c_i \mbox{ , } 0 \leq i < N_{res} \nonumber \\
&&\!\!\!\!\!c_i~e_i \nonumber \\ 
&+& c_i~\int_{0}^{+\infty} \left( u_i(r)^2~V_c(\Delta Z_c,r) - w_{i_s}~s_{k_{i_s}}(r)^2~\frac{C_c~\Delta Z_c}{r} \right) dr \nonumber \\
&+& c_i~\frac{C_c~\Delta Z_c}{\pi} \left[ (k_{max}+k_{i_s}) \ln(k_{max}+k_{i_s}) \right. \nonumber \\
&-& \left. (k_{max}-k_{i_s}) \ln(k_{max}-k_{i_s}) - 2 k_{i_s} \ln(k_{i_s}) \right]  \nonumber \\
&-& c_i~\frac{C_c~\Delta Z_c}{\pi}~\sum_{i'_s=0}^{i_s-1}  w_{i'_s}~[\ln(k_{i_s} + k_{i'_s}) - \ln(k_{i_s} - k_{i'_s})] \nonumber \\
&-& c_i~\frac{C_c~\Delta Z_c}{\pi}~\sum_{i'_s=i_s+1}^{N_{GL}-1}  w_{i'_s}~[\ln(k_{i_s} + k_{i'_s}) - \ln(k_{i'_s} - k_{i_s})] \nonumber \\
&+& \sum_{\stackrel{i'=0}{i' \neq i}}^{N-1} c_{i'} \langle u_{i'} | V_c(\Delta Z_c,r) | u_i \rangle = E~c_i \mbox{, } N_{res} \leq i < N \label{subtraction_discretized_Fredholm_equations}.
\end{eqnarray}

\subsection{The off-diagonal method} \label{off_diagonal_method}
The exchange of the diagonal infinite matrix elements by close off-diagonal matrix elements is motivated by the analytical approximation of the integral of $\ln|k-k'|$ between $k'=0$ and $k'=k_{max}$ 
acquired from the trapezoidal rule.
For that, one uses two grids of equally space points on $[0:k]$ and $[k:k_{max}]$, 
of same discretization step $\Delta k$ ($k_{max}$ is assumed to be an integral multiple of $\Delta k$),
so that $N = k/\Delta k$ and $M = (k_{max}-k)/\Delta k$ are their respective number of points.
Removing the infinite value at $k' = k$ from the sums deduced from the trapezoidal rule, one obtains:
\begin{eqnarray}
&& \int_0^{k_{max}} \ln|k-k'|~dk' = \int_0^k \ln(k-k')~dk' \nonumber \\
&+& \int_k^{k_{max}} \ln(k'-k)~dk' \nonumber \\
&\rightarrow&\sum_{i=0}^{N-1} \ln(k - i \Delta k)~\Delta k + \sum_{i=1}^{M} \ln(i \Delta k)~\Delta k  \nonumber \\
&-& \frac{\Delta k}{2} \ln(k) - \frac{\Delta k}{2} \ln(k_{max}-k)  \nonumber \\
&=& [N \ln(\Delta k) + \ln(N!) + M \ln(\Delta k) + \ln(M!)] \Delta k \nonumber \\
&-& \frac{\Delta k}{2} \ln(k)- \frac{\Delta k}{2} \ln(k_{max}-k) \nonumber \\
&=& (k_{max}-k) \ln(k_{max}-k) - (k_{max}-k)  \nonumber \\
&+& k \ln(k) - k - \Delta k \ln \left( \frac{\Delta k}{2 \pi} \right) + O({\Delta k}^2), \nonumber \\
&&\Delta k \rightarrow 0 \label{trapezoidal_rule}.
\end{eqnarray}
If one replaces the infinite value of $\ln|k-k'|$ at $k' = k$ by $\ln [\Delta k / (2 \pi)]$ in the first line of Eq.~(\ref{trapezoidal_rule}), its end point contribution of the trapezoidal rule 
of the integrals defined on $[0:k]$ and $[k:k_{max}]$ at $k' = k$ is equal to $(\Delta k / 2) \ln[\Delta k/ (2 \pi)]$, so that $\displaystyle \int_0^{k_{max}} \ln|k-k'|~dk' $
is reproduced up to ${\Delta k}^2$ with the trapezoidal rule, as $\Delta k \ln[\Delta k/ (2 \pi)]$ is then added to Eq.~(\ref{trapezoidal_rule}).
As a consequence, we introduce the transformation $k \rightarrow k - \Delta k /(4 \pi)$ and $k' \rightarrow k + \Delta k /(4 \pi)$ at $k = k'$.
This has the advantage not to spoil the precision of the integral if one
adds a symmetric function $f(k,k')$ regular at $k = k'$ to $\ln|k-k'|$,
as $f(k  - \Delta k /(4 \pi),k + \Delta k /(4 \pi)) = f(k,k) + O({\Delta k}^2)$, immediate from the second-order Taylor expansion of $f(k,k')$ at $k = k'$.
Note that $O({\Delta k}^2)$ is the error one expects from the standard trapezoidal rule.

\begin{figure*}[hbt]
\begin{center}
\includegraphics[width=1.0\textwidth,angle=00]{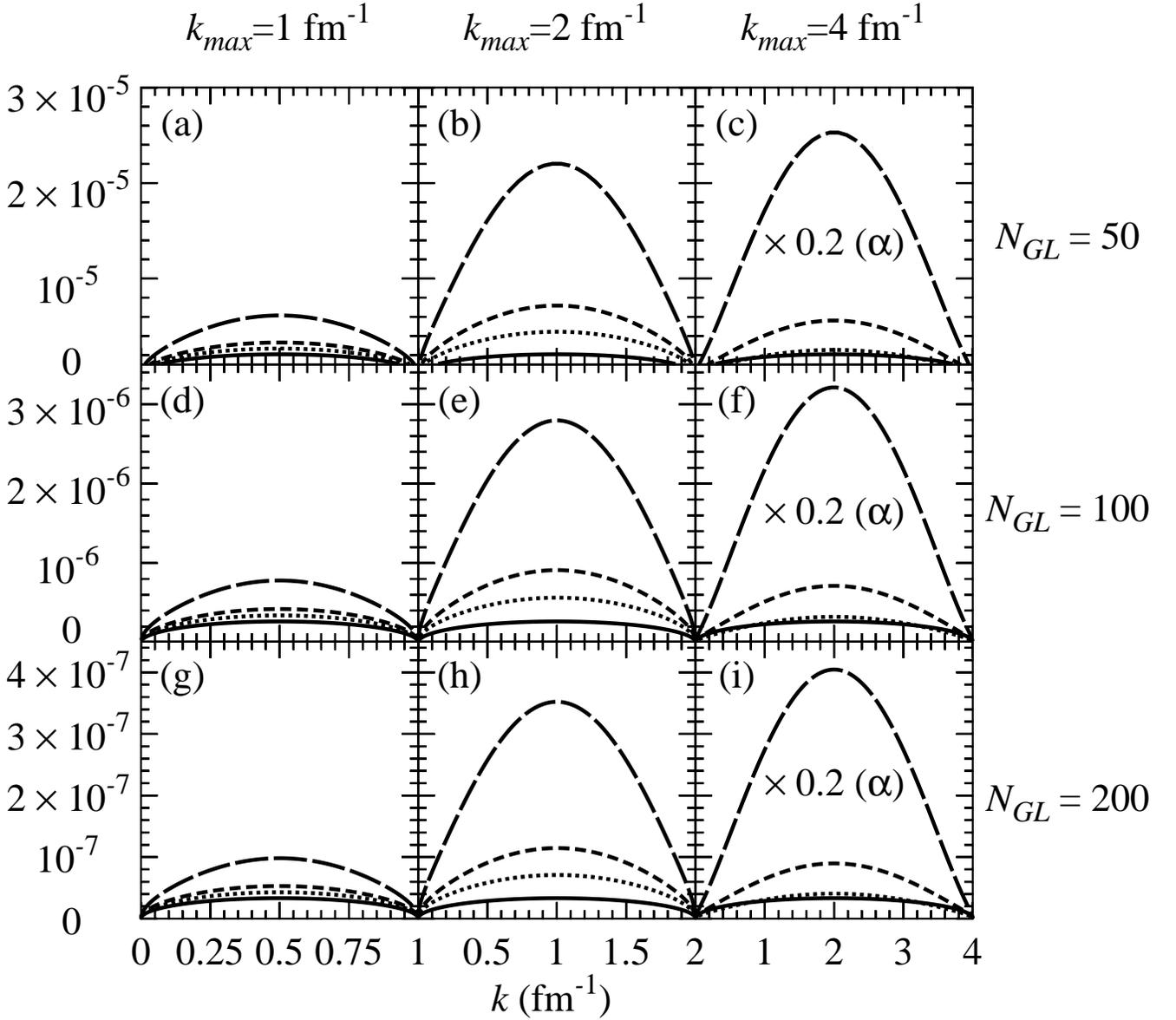}
\caption{Relative difference $\Delta I(N_{GL},k,k_{max})$ [see Eq.~(\ref{relative_difference})] as a function of $k$. 
$N_{GL}$ is the number of Gauss-Legendre points in $I_{GL}(N_{GL},k,k_{max})$ [see Eq.~(\ref{studied_integral_discretized})]
and $k$ varies from $k=0$ to $k=k_{max}$. We consider $N_{GL}=$ 50, 100 and 200, $k_{max}=$ 1, 2 and 4 fm$^{-1}$ and $\alpha=$ 0.25, 0.45 and 0.65 fm$^{-1}$,
where $\alpha$ defines the Coulomb potential $V_c(\Delta Z_c,r)$ used [see Eq.~(\ref{Coulomb_potential})]. 
Results are independent of $\Delta Z_c$ as it simplifies in $\Delta I(N_{GL},k,k_{max})$ [see Eq.~(\ref{relative_difference})].
Long-dashed, dashed, and dotted lines correspond respectively to $\alpha=$ 0.25, 0.45 and 0.65 fm$^{-1}$.
Solid lines refer to the use of a point-particle Coulomb potential
 $(C_c~\Delta Z_c)/r$ instead of $V_c(\Delta Z_c,r)$ in
 Eqs.~(\ref{Coulomb_potential}), (\ref{studied_integral}), and (\ref{studied_integral_discretized}).
Data corresponding to $k_{max}=4$ fm$^{-1}$ and $\alpha=$ 0.25, 0.45 and
 0.65 fm$^{-1}$ have been scaled down by a factor of five, as indicated
 by ``$\times$ 0.2 $(\alpha)$''.}
\label{GL_transformation_numerical_study}
\end{center}
\end{figure*}

Even though one cannot assess analytically the error made using
Gauss-Legendre quadrature instead of the trapezoidal rule,
it will be proved to be very small numerically. For that, we consider the following integral:
\begin{eqnarray}
I(k,k_{max}) = \int_0^{k_{max}} \langle s_{k'} | V_c(\Delta Z_c,r) | s_k \rangle~dk',  \label{studied_integral}
\end{eqnarray}
which is the integral entering Eq.~(\ref{Fredholm_equations}) possessing the same singularities as that of Eq.~(\ref{trapezoidal_rule}), 
where basis functions $u_k(r)$ have been replaced by sine functions $s_k(r)$ and the complex $L^+$ contour is hereby the real segment $[0:k_{max}]$.
If one takes $V_c(\Delta Z_c,r)$ to be exactly equal to the point Coulomb potential $(C_c~\Delta Z_c)/r$ [i.e.~$\alpha \rightarrow +\infty$ in Eq.~(\ref{Coulomb_potential})],
Eq.~(\ref{studied_integral}) is analytical [see Eq.~(\ref{analytical_integral})].
When $\alpha$ is finite, it is nevertheless possible to evaluate almost exactly the integral of Eq.~(\ref{studied_integral}) rewriting $V_c(\Delta Z_c,r)$ as $[V_c(\Delta Z_c,r) - (C_c~\Delta Z_c)/r] + (C_c~\Delta Z_c)/r$: 
\begin{eqnarray}
&& \int_0^{k_{max}} \langle s_{k'} | V_c(\Delta Z_c,r) | s_k \rangle~dk'  \nonumber \\
&=& \int_0^{k_{max}}  \left\langle s_{k'} \left| V_c(\Delta Z_c,r) - \frac{C_c~\Delta Z_c}{r} \right| s_k \right\rangle~dk' \nonumber \\
&+& \frac{C_c~\Delta Z_c}{\pi} \left[ (k_{max}+k) \ln(k_{max}+k) \right. \nonumber \\
&-& \left. (k_{max}-k) \ln(k_{max}-k) - 2 k \ln(k) \right]  \label{studied_analytical_integral}.
\end{eqnarray}
As we consider three finite $\alpha$ values for $V_c(\Delta Z_c,r)$ [see Eq.~(\ref{Coulomb_potential})], namely $\alpha$ = 0.25, 0.45 and 0.65 fm$^{-1}$, 
which correspond respectively to a number of nucleons equal to about 70, 10 and 5,
calculating the radial integrals involving $[V_c(\Delta Z_c,r) - (C_c~\Delta Z_c)/r]$ in Eq.~(\ref{studied_analytical_integral}) with 300 Gauss-Legendre points defined on $[0:R]$, with $R$ = 30 fm, 
guarantees an almost exact reproduction of their numerical value. The Gauss-Legendre approximation of Eq.~(\ref{studied_integral}) then follows from the transformation mentioned above:
\begin{eqnarray}
&&I_{GL}(N_{GL},k_{i_s},k_{max}) = \sum_{\stackrel{i'_s=0}{i'_s \neq i_s}}^{N_{GL}-1} \langle s_{k_{i'_s}} | V_c(\Delta Z_c,r) | s_{k_{i_s}} \rangle~w_{i'_s} \nonumber \\
                              &+& \langle s_{k^+_{i_s}} | V_c(\Delta Z_c,r) | s_{k^-_{i_s}} \rangle~w_{i_s}, \label{studied_integral_discretized} 
\end{eqnarray}
where $0 \leq i_s < N_{GL}$ and $k_{i_s}^\pm$ is defined as:
\begin{eqnarray}
k_{i_s}^\pm = k_{i_s} \pm \frac{w_{i_s}}{4 \pi} \mbox{ , } 0 \leq i_s < N_{GL} \label{k_plus_minus}.
\end{eqnarray}
In order to assess the precision of Eq.~(\ref{studied_integral_discretized}) when compared to Eq.~(\ref{studied_integral}), we define the following relative difference:
\begin{eqnarray}
&&\Delta I(N_{GL},k_{i_s},k_{max})  \nonumber \\
&=& \left| \frac{I_{GL}(N_{GL},k_{i_s},k_{max}) - I(k_{i_s},k_{max})}{\max|I(k_{i'_s},k_{max})|_{0 \leq i'_s < N_{GL}}} \right| \label{relative_difference}.
\end{eqnarray}
$\Delta I(N_{GL},k_{i_s},k_{max})$ is plotted for various values of $N_{GL}$ and $k_{max}$ on Fig.~(\ref{GL_transformation_numerical_study}).
One can see that it is very small, of the order of $10^{-6}$ to $10^{-4}$ for $N_{GL}=50$, 
and decreases by an order of magnitude when $N_{GL}$ is multiplied by a factor of two. 
Interestingly, the shape of $\Delta I(N_{GL},k_{i_s},k_{max})$ remains identical when $N_{GL}$ is modified, with $\alpha$ and $k_{max}$ kept fixed.
Precision is comparable to that of point-particle Coulomb potential for $\alpha=$ 0.45 and 0.65 fm$^{-1}$, the values of interest
for the lightest nuclei, which are the most studied with the Berggren basis \cite{Mic02,Mic03,Mic04,Gaute_Morten,Gaute_Morten_2,Gaute_Morten_3,Mic07,Mic07a,Mic10}.
As $N_{GL}$ rarely exceeds 100 in practical applications and $k_{max}$ is typically of the order of 4 fm$^{-1}$ \cite{Mic10},
the integral discretization inspired from Eq.~(\ref{trapezoidal_rule}) is justified with Gauss-Legendre quadrature.

We can now proceed to the discretization scheme we advocate for the general Berggren basis from a subtraction method similar to that of Eq.~(\ref{subtraction_integral}).
The notations used will be identical to those of Eq.~(\ref{subtraction_discretized_Fredholm_equations}) and Sec.~\ref{Berggren_basis_discretization}.
We will also use the following definition:
\begin{eqnarray}
u^\pm_i(r) = \sqrt{w_{i_s}}~u_{k_{i_s}^\pm}(r)\mbox{ , } N_{res} \leq i < N \label{u_plus_minus},
\end{eqnarray}
where $k_{i_s}^\pm$ is defined in Eq.~(\ref{k_plus_minus}).
The subtraction scheme from which we start is the following:
\begin{eqnarray} 
&&\int_{L^+} c_{k'} \langle u_{k'} | V_c(\Delta Z_c,r) | u_k \rangle~dk' \nonumber \\ 
&=& \int_{L^+} \left[ \int_{0}^{+\infty} \left( c_{k'}~u_{k'}(r)~u_k(r) \right. \right.  \nonumber \\ 
&-& \left. \left. c_k~s_{k'}(r)~s_k(r) \right) V_c(\Delta Z_c,r)~dr \frac{}{} \right] dk' \nonumber \\ 
&+& c_{k} \int_{L^+} \langle s_{k'} | V_c(\Delta Z_c,r) | s_k \rangle~dk' \label{temporary_subtraction_integral},
\end{eqnarray}
where Coulomb point potential is no longer utilized, contrary to that of Eq.~(\ref{subtraction_integral}).
Based on the latter theoretical and numerical arguments, the discretization scheme of Eq.~(\ref{subtraction_discretized_Fredholm_equations}) becomes for Eq.~(\ref{temporary_subtraction_integral}):
\begin{eqnarray}
&&c_i~e_i + \sum_{i'=0}^{N-1} c_{i'} \langle u_{i'} | V_c(\Delta Z_c,r) | u_i \rangle = E~c_i \mbox{ , } 0 \leq i < N_{res} \nonumber \\
&& c_i~\left[ e_i + \int_{0}^{+\infty} \left( u_i(r)^2 - w_{i_s}~s_{k_{i_s}}(r)^2 \right) V_c(\Delta Z_c,r)~dr \right] \nonumber \\
&+& c_i~w_{i_s}~\int_0^{+\infty} s_{k_{i_s}^+}(r) V_c(\Delta Z_c,r) s_{k_{i_s}^-}(r)~dr \nonumber \\
&+& \sum_{\stackrel{i'=0}{i' \neq i}}^{N-1} c_{i'} \langle u_{i'} | V_c(\Delta Z_c,r) | u_i \rangle = E~c_i \mbox{ , } N_{res} \leq i < N \label{temporary_subtraction_discretized_Fredholm_equations}.
\end{eqnarray}
All off-diagonal matrix elements involving $s_{k_{i_s}}(r)$ and $s_{k'_{i_s}}(r)$ issued from the integrals of the right-hand side of Eq.~(\ref{temporary_subtraction_integral}) cancel out.
Indeed, $\displaystyle \int_{0}^{+\infty} \left( c_{k'}~u_{k'}(r)~u_k(r) - c_k~s_{k'}(r)~s_k(r) \right) V_c(\Delta Z_c,r)~dr$ can always be rewritten as
$c_{k'}~\langle u_{k'} | V_c(\Delta Z_c,r) | u_k \rangle - c_k~\langle s_{k'} | V_c(\Delta Z_c,r) | s_k \rangle$, where the two latter matrix elements can be calculated with complex rotation of $r$ as $k \neq k'$ 
(see Sec.~\ref{theory_Berggren_basis}). 
One can verify that replacing $u_i(r)^2$ by $u^+_i(r) u^-_i(r)$ and
$s_{k_{i_s}}(r)^2$ by $s_{k_{i_s}^+}(r) s_{k_{i_s}^-}(r)$ [see Eqs.~(\ref{k_plus_minus}) and (\ref{u_plus_minus})]
in the radial integral of the second line of Eq.~(\ref{temporary_subtraction_discretized_Fredholm_equations}) results in an error of the order of $w_{i_s}^3 / (4 \pi)^2$, which can be neglected. 
Because $k_{i_s}^+ \neq k_{i_s}^-$, the cancellation we have noticed to take place for off-diagonal matrix elements now occurs in the modified diagonal radial integral.
The off-diagonal discretization scheme then reads:
\begin{eqnarray}
&&c_i~e_i + \sum_{i'=0}^{N-1} c_{i'} \langle u_{i'} | V_c(\Delta Z_c,r) | u_i \rangle  = E~c_i \mbox{ , } 0 \leq i < N_{res} \nonumber \\
&&c_i~\left[ e_i + \langle u^+_i | V_c(\Delta Z_c,r) | u^-_i \rangle \right] \nonumber \\
&+& \sum_{\stackrel{i'=0}{i' \neq i}}^{N-1} c_{i'} \langle u_{i'} | V_c(\Delta Z_c,r) | u_i \rangle = E~c_i \mbox{ , } N_{res} \leq i < N \label{off_diagonal_discretized_Fredholm_equations}.
\end{eqnarray}
Note that the $k_{max}$ dependence of Eq.~(\ref{subtraction_discretized_Fredholm_equations}) has disappeared in Eq.~(\ref{off_diagonal_discretized_Fredholm_equations}),
so that this scheme can be used in principle when $k_{max} \rightarrow +\infty$.

\section{Numerical examples} \label{examples} 
We consider now the expansion of the three single-particle states of the $sd$-shell with a Berggren basis, namely proton $1s_{1/2}$, $0d_{5/2}$ and $0d_{3/2}$ wave functions.
The parameters defining the Woods-Saxon potential $V_{WS}(r)$ and
Coulomb potential $V_c(Z_c,r)$ [see Eqs.~(\ref{WS_potential}) and (\ref{Coulomb_potential})], 
common for basis and diagonalized Hamiltonians [see Eq.~(\ref{Hamiltonian})], are $d$ = 0.65 fm, $R_0$ = 3 fm, $V_o$ = 52 MeV, $V_{so}$ = 5 MeV and $\alpha = 3 \sqrt{\pi} / (4 R_0)$. 
The latter value for $\alpha$ is chosen in order for $V_c(Z_c,r=0)$ to sustain the same value as that of the Coulomb potential defined from a uniformly charged-sphere of radius $R_0$ \cite{Mic10}.
Basis and diagonalized Hamiltonians differ through the value of $Z_c$ in Eq.~(\ref{Coulomb_potential}), in which $Z_c^{(b)}$ = 10 and $Z_c^{(d)}$ = 8 (see Sec.~\ref{theory_Berggren_basis} for notations).
These parameters are typical of Woods-Saxon potentials mimicking nuclei of $A \sim 20$ nucleons. 
The energies and widths of proton states of the basis and diagonalized
Hamiltonians, calculated with direct integration, are given in Table \ref{E_Gamma_exact}.
Note that the proton $0s_{1/2}$ state appears in the $s_{1/2}$ Berggren
basis, but, as it is well bound, the proton $0s_{1/2}$ state just bears perturbative modification from basis to diagonalized Hamiltonian,
so that it is not interesting to study its case.
It is important to state that it is not our aim to reproduce experimental data, but only to study how precise Berggren basis expansion is if an infinite-range Coulomb potential is present in the diagonalized residual interaction.
\begin{table*}[htb]
\renewcommand{\arraystretch}{1.2}
\renewcommand\tabcolsep{3pt}
\caption[T3]{Energies and widths of proton  $1s_{1/2}$, $0d_{5/2}$ and $0d_{3/2}$ states for both basis and diagonalized Hamiltonians, determined with direct integration. Energies are given in MeV and widths in keV. The basis states are denoted as ``basis'' and the diagonalized states as ``diag.''}
\label{E_Gamma_exact}
\begin{ruledtabular}
\begin{tabular}{l|llll} 
                      & $E$ basis (MeV)   & $\Gamma$ basis (keV)  & $E$ diag (MeV)  & $\Gamma$ diag (keV)
\\ \hline
$1s_{1/2}$            & 1.09747       & 134.623            & 0.463324    & 8.96828 
\\
$0d_{5/2}$            & 1.48359       & 11.9527            & 0.666208    & 0.525611
\\
$0d_{3/2}$            & 5.07435       & 1353.51            & 4.3003      & 1091.3 
\\  \hline
\end{tabular} 
\end{ruledtabular}
\end{table*}
When wielding the cut method (see Sec.~\ref{cut_method}),
the cut radius $R$ of Eq.~(\ref{Vc_complex_scaling_OBME}) will be fixed
at $R$ = 75 fm for $s_{1/2}$ and $d_{5/2}$ partial waves, and at $R$ =
35 fm for the $d_{3/2}$ partial wave,
which are the values yielding the best precision for the cut method, while the Berggren basis contour [see Eq.~(\ref{Berggren_completeness_relation})] consists of three segments of the complex $k$-plane, 
delimited by the four points $k_{min}$ = 0 fm$^{-1}$, $k$ = 0.25-0.1$i$ fm$^{-1}$ ($s_{1/2}$ and $d_{5/2}$ partial waves) or $k$ = 0.4-0.39$i$ fm$^{-1}$ ($d_{3/2}$ partial wave), $k$ = 1 fm$^{-1}$, and $k_{max}$ = 4 fm$^{-1}$.  

The radial contour employed for integration with complex rotation of $r$ in Eq.~(\ref{Vc_complex_scaling_OBME}) is defined by fixing $R$ = 15 fm and $\theta$ = $-3\pi/4$, $-\pi/4$, $\pi/4$ or $3 \pi/4$.
Improper integrals of Eq.~(\ref{Vc_complex_scaling_OBME}) are indeed guaranteed to converge employing one of these angular values.
The Berggren basis contour used in Eq.~(\ref{Berggren_completeness_relation}) is very similar to that of the cut method, except that $k_{min}$ is chosen so that $|F_{\ell \eta}(k_{min} R)| + |k_{min} F'_{\ell \eta}(k_{min} R)| = 10^{-5}$.
Indeed, it is necessary for $k_{min}$ to be stricly positive, because, on the one hand, proton scattering states are very close to regular Coulomb wave functions $\sqrt{2 / \pi} F_{\ell \eta}(kr)$ when $k \rightarrow 0$ \cite{JMP}, 
very small for moderate values of $r$, and on the other hand, $H^\pm_{\ell \eta}(kr)$ functions enter integration effected with complex rotation of $r$ in Eq.~(\ref{Vc_complex_scaling_OBME}), 
which bear very large modulus for moderate values of $r$.
As a consequence, very important numerical cancellations occur between the different improper integrals of Eq.~(\ref{Vc_complex_scaling_OBME}). Nevertheless, proton scattering states
with very small linear momentum play virtually no role in Berggren basis completeness, because of the extreme smallness of their amplitudes.
Hence, proton scattering states become important only when $k_{min}$ become sufficiently large, and the condition above has been shown to mitigate the numerical instability occurring for $k \sim 0$ while yielding precise results.

\begin{figure*}[hbt]
\begin{center}
\includegraphics[width=1.0\textwidth,angle=00]{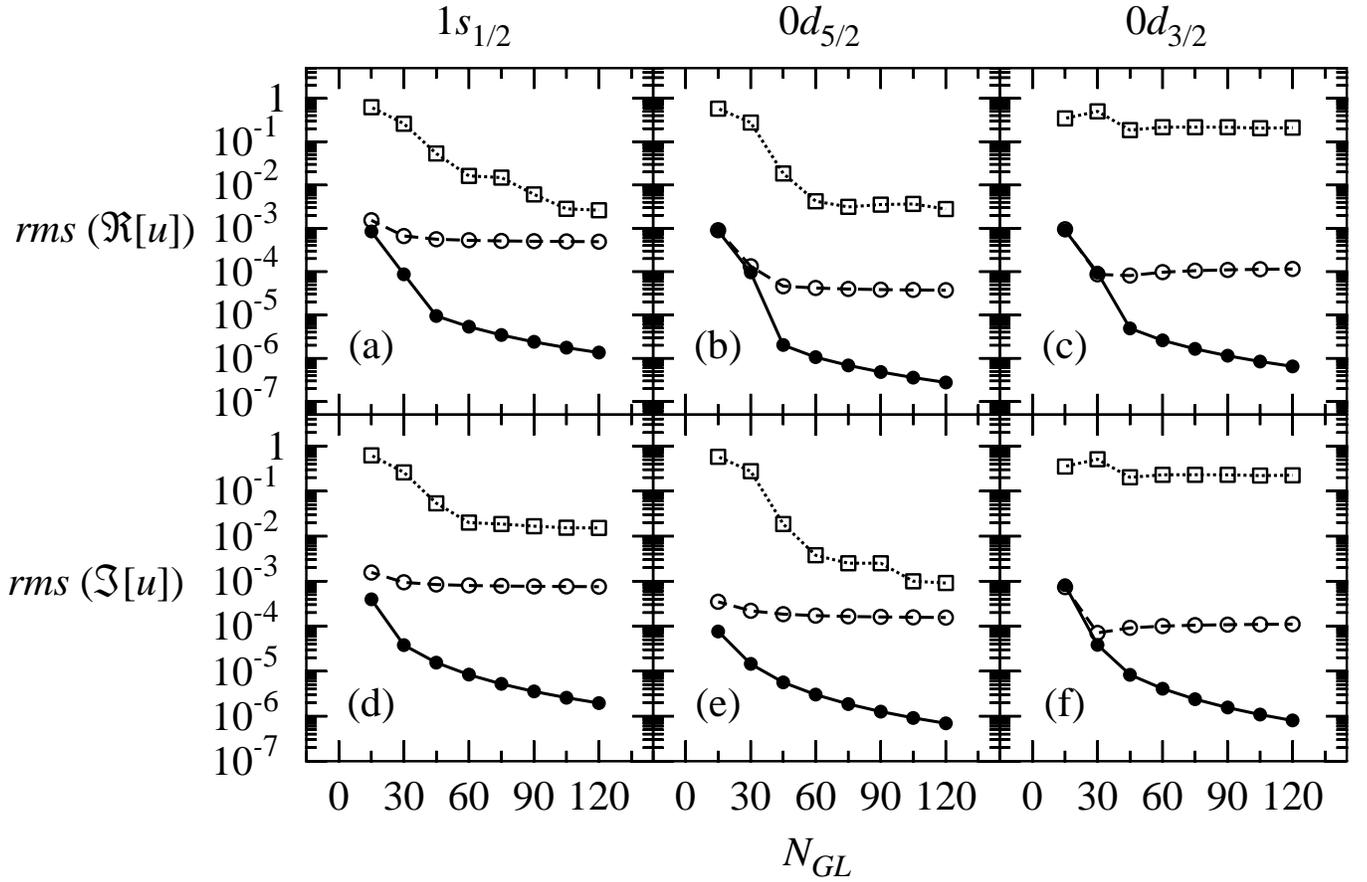}
\caption{Root mean squares of the real [$rms(\Re[u])$] and imaginary
 [$rms(\Im[u])$] parts of diagonalized wave functions $u(r)$ of proton
 $1s_{1/2}$, $0d_{5/2}$ and $0d_{3/2}$ states [see Eqs.~(\ref{rms_Re}) and (\ref{rms_Im})],
as a function of the number of scattering states discretized with Gauss-Legendre quadrature $N_{GL}$. 
Dotted lines with empty squares refer to the cut method (see Sec.~\ref{cut_method}), 
dashed lines with empty circles to the subtraction method (see Sec.~\ref{subtraction_method}), 
and solid lines with filled circles to the off-diagonal method (see Sec.~\ref{off_diagonal_method}).
Symbols depict the $N_{GL}$ values utilized in calculations while lines are to guide the eyes.}
\label{rms_Re_Im_u}
\end{center}
\end{figure*}

Comparison of energies and widths of considered proton states provided
by diagonalization to those deduced from direct integration are depicted
in Tables \ref{E_Gamma_1s1I2}, \ref{E_Gamma_0d5I2}, and \ref{E_Gamma_0d3I2}.
\begin{table*}[htb]
\renewcommand{\arraystretch}{1.2}
\renewcommand\tabcolsep{3pt}
\caption[T3]{Energies and widths of proton $1s_{1/2}$ state issued from the diagonalization of the Hamiltonian of Eq.~(\ref{Hamiltonian}) with $Z_c^{(d)}$ = 8,
expanded from a Berggren basis generated by a Hamiltonian of the same structure but bearing $Z_c^{(b)}$ = 10 
[see Eqs.~(\ref{WS_potential}) and (\ref{Coulomb_potential}) and Sec.~\ref{examples} for the values of the other Hamiltonian parameters], 
as a function of the number of scattering states $N_{GL}$ of the $L^+$ contour of Eq.~(\ref{Berggren_completeness_relation}), discretized with Gauss-Legendre quadrature.
$E$ designates the energy of the proton state, given in MeV, and $\Gamma$ the width of the proton state, given in keV, while ``exact'' refers to results obtained with direct integration.
The different discretization methods [i.e. the cut method, the
 subtraction method and the off-diagonal method (see Secs.~\ref{cut_method}, \ref{subtraction_method}, and \ref{off_diagonal_method})],
are denoted respectively as ``cut,'' ``sub,'' and ``off-diag.''}
\label{E_Gamma_1s1I2}
\begin{ruledtabular}
\begin{tabular}{l|llllll} 
$N_{GL}$         & $E$ cut (MeV)   & $\Gamma$ cut (keV)  & $E$ sub (MeV)  & $\Gamma$ sub (keV)  & $E$ off-diag (MeV)  & $\Gamma$ off-diag (keV)  
\\ \hline
15              & 0.461875    & -11.6596        & 0.464574    & 9.19011        & 0.46396         & 10.2211
\\
30              & 0.465707    & 13.4833         & 0.463777    & 8.26812        & 0.463343        & 8.97219
\\
45              & 0.463476    & 8.71097         & 0.463709    & 8.33267        & 0.463334        & 8.96171
\\
60              &  0.463307   & 8.68396         & 0.463681    & 8.36454        & 0.463329        & 8.96458
\\ 
75              &  0.463227   & 8.70558         & 0.463667    & 8.38006        & 0.463328        & 8.96595
\\
90              &  0.46284    & 8.88896         & 0.463659    & 8.3888         & 0.463327        & 8.96669
\\
105             &  0.462952   & 8.69106         & 0.463654    & 8.39421        & 0.463326        & 8.96712
\\
120             &  0.462949   & 8.62468         & 0.46365     & 8.3978         & 0.463326        & 8.9674
\\  \hline
exact           &  0.463324   & 8.96828         & 0.463324    & 8.96828        & 0.463324        & 8.96828
\\  \hline
\end{tabular}
\end{ruledtabular} 
\end{table*}
\begin{table*}[htb]
\renewcommand{\arraystretch}{1.2}
\renewcommand\tabcolsep{3pt}
\caption[T3]{Same as Table \ref{E_Gamma_1s1I2}, but for the proton $0d_{5/2}$ state.}
\label{E_Gamma_0d5I2}
\begin{ruledtabular}
\begin{tabular}{l|llllll} 
$N_{GL}$         & $E$ cut (MeV)   & $\Gamma$ cut (keV)  & $E$ sub (MeV)  & $\Gamma$ sub (keV)  & $E$ off-diag (MeV)  & $\Gamma$ off-diag (keV)  
\\ \hline
15              & 0.664431    & 3.68108         & 0.666482   & 0.234635        & 0.666428        & 0.174831    
\\
30              & 0.665162    & -6.30699        & 0.666251   & 0.586056        & 0.666216        & 0.525826
\\
45              & 0.66623     & 0.502731        & 0.66624    & 0.578586        & 0.666209        & 0.527655
\\
60              & 0.666207    & 0.52976         & 0.666237   & 0.573284        & 0.666209        & 0.526746
\\ 
75              & 0.666209    & 0.536367        & 0.666236   & 0.5705          & 0.666209        & 0.526332 
\\
90              & 0.666206    & 0.484164        & 0.666235   & 0.568849        & 0.666209        & 0.52611 
\\
105             & 0.666225    & 0.502034        & 0.666234   & 0.567786        & 0.666209        & 0.525978
\\
120             & 0.666226    & 0.539324        & 0.666234   & 0.567058        & 0.666209        & 0.525892
\\  \hline
exact           & 0.666208    & 0.525611        & 0.666208   & 0.525611        & 0.666208        & 0.525611  
\\  \hline
\end{tabular} 
\end{ruledtabular}
\end{table*}
\begin{table*}[htb]
\renewcommand{\arraystretch}{1.2}
\renewcommand\tabcolsep{3pt}
\caption[T3]{Same as Table \ref{E_Gamma_1s1I2}, but for the proton $0d_{3/2}$ state.}
\label{E_Gamma_0d3I2}
\begin{ruledtabular}
\begin{tabular}{l|llllll} 
$N_{GL}$         & $E$ cut (MeV)   & $\Gamma$ cut (keV)  & $E$ sub (MeV)  & $\Gamma$ sub (keV)  & $E$ off-diag (MeV)  & $\Gamma$ off-diag (keV)  
\\ \hline
15              & 4.29692     & 1091.1          & 4.30016    & 1091.48         & 4.30017         & 1091.49
\\
30              & 4.30301     & 1082.24         & 4.3003     & 1091.29         & 4.30031         & 1091.3
\\
45              & 4.30016     & 1091.42         & 4.3003     & 1091.29         & 4.3003          & 1091.3
\\
60              & 4.30064     & 1091.92         & 4.3003     & 1091.3          & 4.3003          & 1091.3
\\
75              & 4.30072     & 1092.15         & 4.3003     & 1091.3          & 4.3003          & 1091.3
\\
90              & 4.30064     & 1091.93         & 4.3003     & 1091.3          & 4.3003          & 1091.3
\\
105             & 4.30069     & 1091.41         & 4.3003     & 1091.3          & 4.3003          & 1091.3
\\
120             & 4.30053     & 1091.58         & 4.3003     & 1091.3          & 4.3003          & 1091.3
\\  \hline
exact           & 4.3003      & 1091.3          & 4.3003     & 1091.3          & 4.3003          & 1091.3
\\  \hline
\end{tabular}
\end{ruledtabular} 
\end{table*}
Precision of radial wave functions has also been taken into account by calculating the root mean squares of their real and imaginary parts, defined by:
\begin{eqnarray}
rms (\Re[u]) &=& \sqrt{\frac{\displaystyle \sum_{i=1}^N \left( \Re[u(r_i)] - \Re[u_e(r_i)] \right)^2}{N}} \label{rms_Re} \\
rms (\Im[u]) &=& \sqrt{\frac{\displaystyle \sum_{i=1}^N \left( \Im[u(r_i)] - \Im[u_e(r_i)] \right)^2}{N}} \label{rms_Im} 
\end{eqnarray}
where $N$ is taken equal to 512, $r_i = i \cdot (R/N)$, for $1 \leq i \leq N$, is a set of uniformly distributed radii of $[0:R]$, 
$u(r)$ is the diagonalized wave function, and $u_e(r)$ is the exact wave function, issued from direct integration.
Root mean squares are illustrated in Fig.~\ref{rms_Re_Im_u} for all studied cases, 
(i.e.~the cut method, the subtraction method and the off-diagonal method; see Secs.~\ref{cut_method}, \ref{subtraction_method}, and \ref{off_diagonal_method}).
One can see that the cut method produces very poor results, 
as it does not even reach the precision acquired with the off-diagonal method when the smallest value of the number of Gauss-Legendre scattering states $N_{GL}$ (see Sec.~\ref{Berggren_basis_discretization}) is employed. 
Moreover, for the $0d_{3/2}$ proton state, although a rather good description of energy and width occurs, 
the reproduction of the wave function is mediocre. Added to that, the choice of the cut radius could only be effected by comparison with exact results,
whereas it has been checked that the two other methods are very robust when the $L^+$ contour parameters are changed. 
The best reproduction of the considered proton states clearly arises with the off-diagonal method, for energies, widths and wave functions.
The subtraction method, while not being completely inaccurate, saturates very quickly to a wrong value for energies, widths and wave functions, when $N_{GL}$ increases.
On the contrary, exponential convergence occurs with the off-diagonal method for both real and imaginary parts of the wave function when $N_{GL}$ augments.
This is an intriguing phenomenon, as the subtraction scheme of Eq.~(\ref{subtraction_discretized_Fredholm_equations}) is based on an exact calculation of the integral exhibiting singularities, leaving a well-defined function
to be integrated numerically, whereas that of Eq.~(\ref{off_diagonal_discretized_Fredholm_equations}), which could be expected at best to be comparable to the subtraction method 
(see Sec.~\ref{subtraction_method} and \ref{off_diagonal_method}), surprisingly surpasses the latter by several orders of magnitudes (see Fig.~\ref{rms_Re_Im_u})). 
It can be explained by noticing that the integrand of the first integral of the right-hand side of Eq.~(\ref{subtraction_integral}), 
constituted by radial integrals converging with complex rotation of $r$ [see Eq.~(\ref{Vc_complex_scaling_OBME})],
while everywhere finite, is not analytic at $k' \sim k$. Indeed, it is equivalent, up to an unimportant constant, to $(k-k') \ln(k-k')$, which does not even possess a finite derivative with respect to $k'$ at $k' = k$. 
This implies that a Gauss-Legendre discretization of this integral will be far less precise than that of analytic functions, which can usually be well approximated by polynomials. 
On the contrary, the use of $\langle u^+_i | V_c(\Delta Z_c,r) | u^-_i \rangle$ in Eq.~(\ref{off_diagonal_discretized_Fredholm_equations}) 
effectively replaces in Eq.~(\ref{Fredholm_equations}) the function of $k'$ $\langle u_{k'} | V_c(\Delta Z_c,r) | u_{k} \rangle$, singular at $k' = k$, by an analytic function.
Gauss-Legendre quadrature then yields fast convergence to the numerical value of the integral defined by the off-diagonal method, 
which happens to be almost equal to the exact singular kernel (see Sec.~\ref{off_diagonal_method}).

\section{Conclusion} \label{conclusion}
Bases carrying a continuous part, such as Berggren bases, are very
interesting as they allow the expansion of complex many-body wave functions of loosely
bound and unbound states,
as both proper asymptotic properties and particle intercorrelations via configuration mixing are present therein. 
However, the use of Berggren bases is accompanied by mathematical
difficulties generated by the unbound character of the considered one-body states. 
Indeed, their completeness properties request many more efforts to be proved than for discrete sets of states, to which the whole apparatus of the theory of compact operators can be applied.
Furthermore, the very question of their numerical implementation is challenging in the context of charged particles, as Coulomb wave functions have to be calculated for this type of problem.
Discretization of Berggren bases is also central in numerical applications, which is conveniently effected with Gauss quadrature but demands discretization error to be assessed.
Handling of infinite-range interactions is therefore problematic with unbound bases, because the representation of the former give rise to operators bearing singularities, which introduce infinities when Berggren bases are discretized.

In the case of Coulomb potential, kernels are singular but integrable, so that frameworks built from subtraction techniques, 
based on analytical integration of singularities, have been devised in the context of Fourier-Bessel transform.
However, they could not be directly applied to Berggren bases, because they rely on the analytical character of the Fourier-Bessel transform of Coulomb point potential.
Hence, in this paper, three different discretization schemes have been
studied when the Coulomb potential is included in the Hamiltonian to
diagonalize, namely the cut method, 
where the Coulomb potential is suppressed after a finite radius $R$,
the subtraction method, similar to those utilized with Fourier-Bessel transform, and a new framework, the off-diagonal method,
which amounts to substituting diagonal infinite matrix elements by close but finite off-diagonal matrix elements.

Numerical applications have been considered for three different partial waves, with the examples of $1s_{1/2}$, $0d_{5/2}$ and $0d_{3/2}$ resonant proton states, arising typically from studies of nuclei having $A \sim 20$ nucleons.
Woods-Saxon potentials carrying different charges have been employed for basis-generating and diagonalized Hamiltonians.
While the cut method could be expected to convey poor precision, the off-diagonal method has been demonstrated to outperform the standard subtraction method.
This has been explained by the fact that Gauss-Legendre quadrature is applied within the subtraction method to integrate finite but nonanalytic functions, 
whereas only smooth integrands are treated with Gauss-Legendre quadrature within the off-diagonal method. 
According to the present study, the latter technique should be taken into account seriously when diagonalizing Hamiltonians possessing an infinite-range Coulomb part.

\newpage

\section{Acknowledgments}
This work was supported by the Academy of Finland and University of
Jyv{\"a}skyl{\"a} within the FIDIPRO programme.

\end{document}